\documentclass{ws-mpla}

\newcommand{\tot}{\theta_{13}}
\newcommand{\nuebar}{$\bar{\nu}_e\;$}

\usepackage{epsfig}
\usepackage{subfigure}
\usepackage{amsmath}
\usepackage{amsfonts}
\usepackage{amssymb}
\usepackage{longtable}
\usepackage{graphicx}%
\usepackage{epsf}
\usepackage{graphics}
\usepackage{booktabs}

\begin{document}

\markboth{C. Mariani}
{Review of Reactor Neutrino Oscillation Experiments}

\catchline{}{}{}{}{}
%
\title{Review of Reactor Neutrino Oscillation Experiments}
\author{\footnotesize C. Mariani\footnote{mariani@nevis.columbia.edu}}

\address{Department of Physics, Columbia University, New York, NY 10027, USA}

\maketitle
%
%
\begin{abstract}
In this document we will review the current status of reactor neutrino oscillation experiments and present their physics potentials for measuring the $\theta_{13}$ neutrino mixing angle. The neutrino mixing angle $\theta_{13}$ is currently a high-priority topic in the field of neutrino physics. There are currently three different reactor neutrino experiments, \textsc{Double Chooz}, \textsc{Daya Bay} and \textsc{Reno} and a few accelerator neutrino experiments searching for neutrino oscillations induced by this angle. A description of the reactor experiments searching for a non-zero value of $\theta_{13}$ is given, along with a discussion of the sensitivities that these experiments can reach in the near future.
\end{abstract}
\ccode{14.60.Pq,13.15.+g,25.30.Pt,95.55.Vj,28.41.Ak}
%
%
%
\section{Introduction}
\begin{figure}[htbn!]
\centerline{
	\includegraphics[width=0.7\columnwidth]{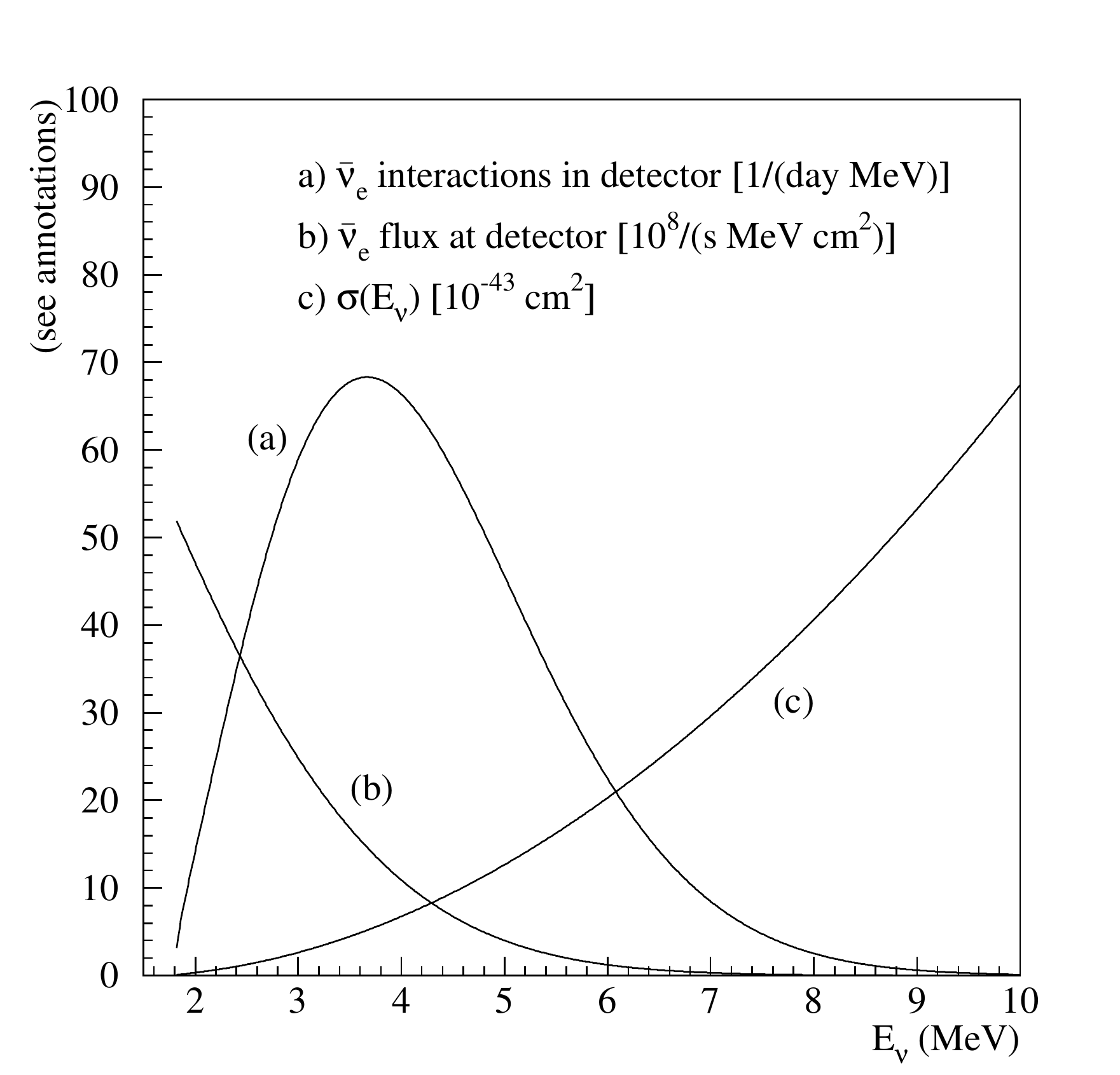}}
\caption{(a) \nuebar interaction in the detector, (b) \nuebar flux and (c) IBD cross section. The Y-axis is in arbitrary unit. Figure from Gratta {\it et al.}\protect\cite{Gratta_paper}}
\label{fig:flux_and_xsec}
\end{figure}

After the observation in recent years of solar\cite{homestake,Fukuda:1996sz,gallex,sage} and atmospheric\cite{Fukuda:1998fd,sno,Fukuda:1998mi} neutrino oscillations and their confirmation at reactor\cite{Chooz,kamland} and accelerator\cite{Ahn:2006zz,Adamson:2011qu,Abe:2011sj} neutrino experiments, the high priority of current neutrino experiments is to measure the $\theta_{13}$ mixing angle, and to improve the accuracy in the measurement of oscillation parameters. A non-zero value for the $\theta_{13}$ is fundamental for the search for CP violation. Reactor experiments play an important role in this search, relying on their unique ability to measure the $\theta_{13}$ angle without coupling with the CP violation phase. The inverse-$\beta$ reaction [IBD] ($\bar{\nu_{e}} + p \rightarrow e^{+} + n$) is one of the oscillations sought for $\theta_{13}$ at reactor neutrino oscillation experiments.  It provides a very good sample of signal events with very little background contamination leading to an excellent signal to background ratio with an energy threshold of 1.8~MeV. An IBD event is identified by looking for a prompt energy deposition followed by a $n$ capture energy signal. These two signals are correlated in time. The \nuebar energy can then be calculated using the deposited energy of the prompt events. Figure~\ref{fig:flux_and_xsec} shows the measured reactor neutrino spectrum with the \nuebar flux from a nuclear reactor and the IBD cross-section\cite{Gratta_paper}. A reactor neutrino detector consists of a central detector filled with liquid scintillator loaded with Gd. The Gd is mixed with the scintillator in order to increase the neutron capture probability and because the neutron capture on Gd has a characteristic energy and capture time that can be used to substantially reduce background events. Multiple gamma-rays are released during the neutron capture on Gd, with a total deposited energy of $\sim$8~MeV, which gives another easy way to isolate signal events and reduce background. In order to improve our knowledge of $\theta_{13}$, high precision experiments with dedicated detector design are required. The statistical and systematic errors should be maintained at less than a 1\% level to improve the limits coming from previous experiments like CHOOZ\cite{CHOOZ}). There are several methods to accomplish this: Use a near detector to reduce rate and spectral shape uncertainties, use large or multiple detectors to increase the fiducial volume and so decrease the statistical error, or use multi-core sites to get as many \nuebar as possible. This should be coupled with an understanding and reduction of the background: deep underground detector sites will help to reduce all the cosmic muon backgrounds and an accurate calibration system will help to reduce background due to radioactivity and systematic uncertainties due to energy reconstruction.
In this paper we discuss the potential of current reactor neutrino experiments to limit or measure $\theta_{13}$. We will analyze the physics potential of the following reactor neutrino experiments: \textsc{Daya Bay}\cite{DayaBay}, \textsc{Double Chooz}\cite{DoubleChooz} and \textsc{Reno}\cite{Reno}. 
%
%
\section{Neutrino Oscillations Framework}\label{sec:oscillation}
The full reactor experiment oscillation probability is reported in Eq.~\ref{eq:oscillation_prob}.
\begin{eqnarray}
1-P_{\bar{e}\bar{e}} \, \simeq \, \sin^22\theta_{13} \, \sin^2\Delta_{31} \, + \,
\alpha^2 \, \Delta_{31}^2  \cos^4\theta_{13} \, \sin^22\theta_{12},
\label{eq:oscillation_prob}
\end{eqnarray}
\noindent where $\Delta_{31} = \Delta m^2_{31} L / 4E$, $L$ is the baseline, and $E$ the neutrino energy. The second term in Eq.~\ref{eq:oscillation_prob}) is important only for reactor experiment with a baseline of the order of 100~km, like Kamland\cite{kamland}.
In this review all calculations are conducted presuming a fixed value of the mass splitting $\Delta m^2_{atm} = 2.32 \times 10^{-3}$ eV$^{2}$, as measured at very high precision by the MINOS experiment\cite{minos_2011}. 
\par Our evaluation of experimental capabilities are performed with a custom sensitivity toolkit, written in C++ with ROOT, which allows a proper treatment of all reactor and detector systematic uncertainties relevant to multi-core and multi-detector experiments. Calculations are executed by generating a set of experimental data with a presumption of no oscillations. This null-oscillation observation is then fit to a predicted data set with a ``pulls approach"\cite{Pulls} multi-parameter minimization, and the uncertainty interval on $\sin^2(2 \tot )$ is evaluated. We have verified our sensitivity code reproducing the results obtained by the Double Chooz experiment: $sin^22\theta_{13} =$ 0.086 $\pm$ 0.041~({\rm stat}) $\pm$0.030~({\rm syst})\cite{DoubleChooz}. A detailed description of all parameters can be found in Sec.~\ref{sec:parameters}.
%
%
%
\section{Experimental Setups}\label{sec:expsetup}
The basic set-up of all the existing (or under construction) reactor experiments consists of a identical near and a far detector. The far detector is generally positioned deeper underground ($\sim$300~m.w.e.) than the near detector ($\sim$150~m.w.e.) to allow more natural shielding from cosmic muons. For the same reason, the Outer Veto detector is bigger for the near than for the far detector. The near detector is typically at an average distance of $\sim$200/300~m from the nuclear reactor cores and the far detector is at $\sim$1000~m.  The nuclear reactors used by the reactor experiments are pressurized water reactors (PWR) and their number and distances with respect to the near detectors varies between experiments\cite{DayaBay,DoubleChooz,Reno}. 
Double Chooz\cite{DoubleChooz} is comprised of the simplest site configuration: two PWR reactors achieving 4.25 GW$_{th}$ each, and two detectors of 8.3 tons fiducial mass each, located in the Ardennes region of France.  The Double Chooz near and far detector sites, situated at respective mean baselines of 395~m and 1050~m, are not placed in a symmetric configuration with respect to the reactor cores, but along a curve of iso-flux such that the ratio of neutrino events from each reactor is equal in the two detectors.  The RENO experiment\cite{Reno} is located near the Yonggwang power station in South Korea, which hosts six reactors of 2.73 GW$_{th}$ power each.  The RENO near and far detectors are located along a line running perpendicular to, and through the midpoint of, the line of reactors, at distances of 290~m and 1,380~m.  Daya Bay's experimental configuration\cite{DayaBay} is situated amongst three separate sites, each hosting two reactors of 2.95 GW$_{th}$ power, in Southeastern China.  Daya Bay's eight detector modules of 20 fiducial tons each will be distributed among three sites. Four detectors will be placed at a ``Far" site, and two each at near sites named for the reactor complexes to which they are closest: Daya Bay (DYB) and Ling Ao (LA).  Daya Bay's Far sites lie between 1,600 and 1,900~m from the sites' reactors, while the near detectors are at baselines of 360~m (for DYB) and 480 to 525~m (for LA, with respect to the Ling Ao and Ling Ao-II reactor sites) to their eponymous reactors.  It is notable that the Daya Bay experiment has two near detector sites while Double Chooz and Reno have only one.
\par
\begin{figure}[htbn!]
\begin{center}
\subfigure[Daya Bay detector\protect\cite{DayaBay}]{\includegraphics[width=0.45\columnwidth]{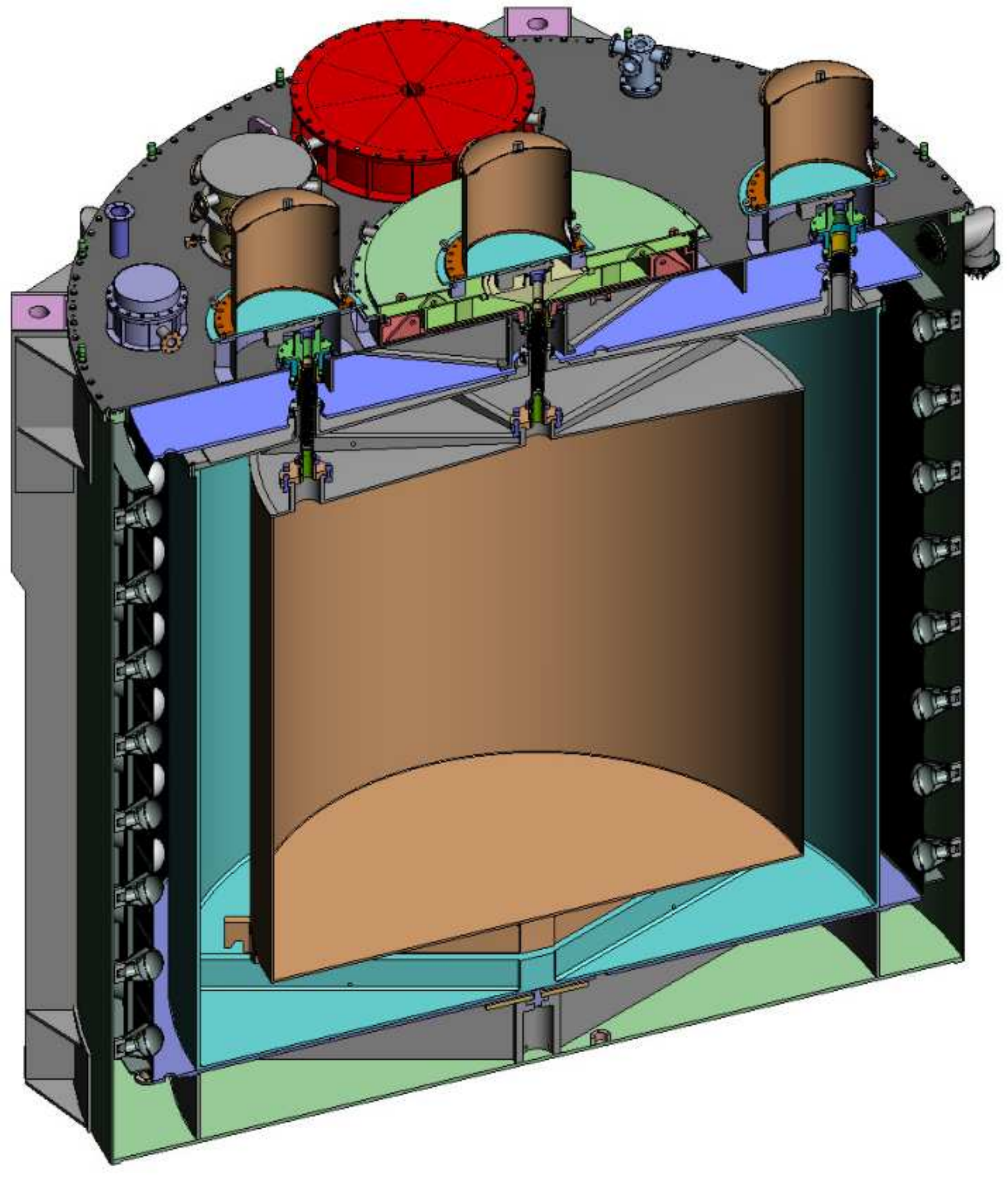}}
\subfigure[Double Chooz Far detector\protect\cite{DoubleChooz}]{\includegraphics[width=0.45\columnwidth]{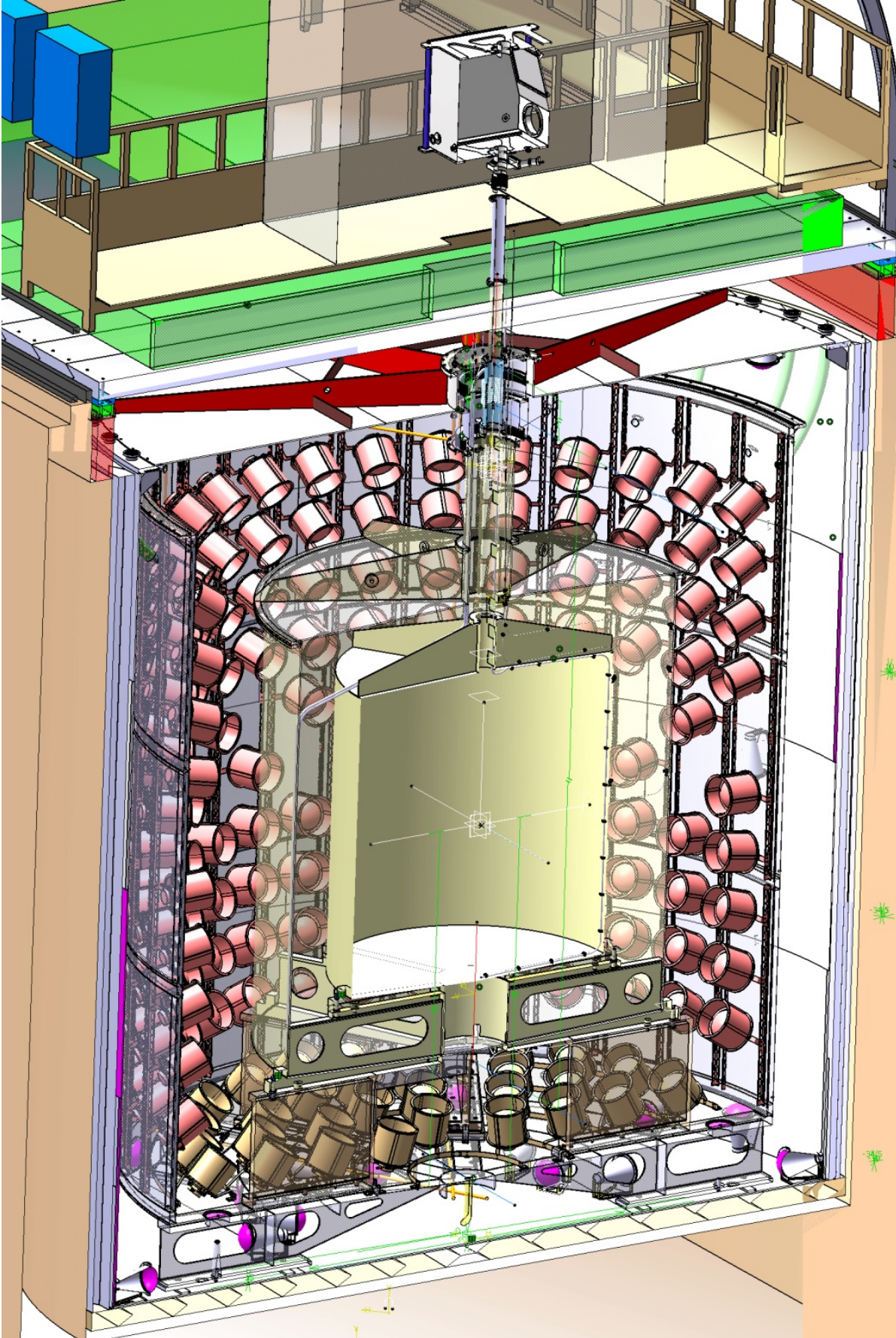}}
\caption{Example of reactor neutrino detectors - cross-sectional view of the Daya Bay\protect\cite{DayaBay} (on the left) and Double Chooz\protect\cite{DoubleChooz} (on the right) detectors.}
  \label{fig:Det}
\end{center} 
\end{figure}

 The near and far detectors are designed to have the same exact fiducial mass of 10 or 20~tons of liquid scintillator with a main detector usually sub-divided into three volumes, an inner veto and an outer veto system. The innermost detector is an acrylic vessel (thickness $\sim$10~mm) that houses the $\nu$-target liquid: a mixture of liquid scintillator doped with Gadolinium. The presence of Gadolinium enhances neutron capture. The design and installation of this volume emphasizes radio-purity and long-term scintillator stability. The $\nu$-target volume is surrounded by a $\gamma$-catcher, a thick ($\sim$60~cm) Gd-free liquid scintillator layer in a second acrylic vessel, used to detect \mbox{$\gamma$-rays} escaping from the $\nu$-target. The light yield of the $\gamma$-catcher scintillator is matched to the $\nu$-target scintillator to provide identical light yield across the full inner volume. Outside the $\gamma$-catcher is the buffer, a $\sim$100~cm thick layer filled with mineral oil. The buffer works to shield  \mbox{$\gamma$-rays} from the radioactivity of photomultipliers (PMTs) and from the surrounding rock. PMTs are installed on the stainless steel buffer tank inner wall to collect light from the inner volumes. These three volumes and the PMTs constitute the inner detector (ID). Outside the ID, and optically separated from it, is a $\sim$50~cm thick ``inner veto'' liquid scintillator (IV). The IV is equipped with generally smaller PMTs and functions as a cosmic muon veto and as a shield to spallation neutrons produced outside the detector. The whole detector is further surrounded by 15~cm of demagnetized steel or a water pool equipped with PMTs as well, to suppress external $\gamma$-rays.  The main detector is covered by an outer veto system (OV) made either of plastic scintillator strips or a system of RPC modules. The OV serves to veto against muons which do not pass through the main detector, but which could act as precursors to cosmogenic backgrounds. The experiment is calibrated using light sources, radioactive point sources, cosmic rays, and natural radioactivity. Deployments in the target are carried out as well using motorized systems and using the glovebox position on the top of the detector. We show the Double Chooz far and the Daya Bay detectors as examples of reactor neutrino detectors in Fig.~\ref{fig:Det}.
%
%
\section{Reactor vs. Accelerator Experiments}\label{sec:accelerator}
Reactor neutrino experiments take advantage of the isotropic emission of \nuebar from the reactor cores, the large IBD cross section and the particular signature of the IBD events in a Gd loaded scintillator detector that helps substantially improving the signal to background ratio.
A reactor-based experiment is a disappearance experiment since the anti-neutrinos produced by the reactor do not have enough energy to produce muons (or taus) and the neutral-current reactions of the ``oscillated'' $\bar\nu_{\mu}$ or $\bar\nu_{\tau}$ or \nuebar NC interactions have very low cross-section values and are indistinguishable from the backgrounds. The three neutrino reactor experiments, \textsc{Daya Bay}, \textsc{Double Chooz} and \textsc{Reno}, which are currently under construction or taking data, have been described in detail in Sec.~\ref{sec:expsetup}. For comparison, neutrino accelerator experiments like \textsc{Minos}\cite{minos} at Fermilab and T2K\cite{t2k} in Japan, will access similar $\Delta m^2$ values with GeV-energy neutrinos and a baseline on the order of 800/900~km. Neutrinos from reactors have a mean energy of 3~MeV, much lower than the energy of the neutrinos produced by accelerator experiments, which means reactor experiments must optimize background reduction.
An important difference between reactor and accelerator experiments is that the oscillation probability for reactor-based neutrino oscillation experiment depends on fewer parameters. The non-dependence upon the $\delta$ CP violating phase and mass hierarchy makes the reactor neutrino experiment the ``cleanest`` and most direct way to measure the neutrino mixing angle $\theta_{13}$. Therefore reactor experiments cannot measure $\delta$ CP violating phase and/or mass hierarchy.
While in the case of neutrinos produced by accelerators, the experiment has full control over the beam (for example when the $\nu$ are produced in time and their directions), the flux of \nuebar from a nuclear reactor cannot be changed. However, in practice, a PWR get refueled every 12 months, this allow the experiment to have a period of data taking with a different neutrino flux and that is used to validate background estimations. A special case here is the Double Chooz experiment that having only two reactor cores have a very good probability to have short periods of time with both reactors off. These periods are extremely valuable because allow the experiment to have a precise background measurements.
%
%
\section{Systematic errors for a reactor neutrino oscillation experiment}\label{sec:systematic}
The sensitivity (or capability) of a reactor neutrino experiment to measure $\theta_{13}$ depends on different factors. Any indication of neutrino oscillations could be due either to the neutrino oscillations or to an incorrect prediction of the anti-neutrino flux from the reactor cores. In past experiments, the determination of the neutrino flux was based on measurement of the reactor thermal power and flux calculation, which nowadays can be done with high precision. However, in current reactor experiments this will no longer be an issue because of the presence of a near detector, closer to the reactor cores, dedicated to the precise measurement of the un-oscillated neutrino flux. The near detector reduces also other uncertainties like ``spill-in/spill-out". An event is define as ``Spill in/out" when the neutrino interaction happen in the neutrino detector target volume, but the neutron escapes and captures into the $\gamma$-catcher or vice-versa. 
\par A special case is the first result from the Double Chooz experiment: the final oscillation was produced using the Far Detector only\cite{DoubleChooz}. In this case only the Far detector was used for the analysis and that required an accurate determination of the \nuebar spectrum and its absolute normalization\cite{DoubleChooz}. We will briefly discuss how this accuracy in the \nuebar spectrum is achieved in Sec.~\ref{sec:reactor_spectrum} for a more accurate discussion please see P. Huber {\it et al.}\cite{Huber:2003pm}.
\par The systematic errors need to be understood at percent level since what a reactor neutrino oscillation experiment aims to detect is a small oscillation disappearance signal.
%
%
\subsection{Backgrounds}\label{sec:backgrounds}
Backgrounds arise from processes mimicking a time coincidence with n-Gd-capture-like emission.
It is very important to maintain backgrounds associated with cosmic muons as low as possible, for this reason the use of different high efficiency Veto systems (like Inner-Veto and Outer-Veto combinations) and build detector under natural hill is very important. Natural hill offers a good detector shield against cosmic ray muons induced backgrounds. To be noticed here that the near and far detectors will have in some cases different background (due to different overburden) and signal (distance to cores) rates. In the following we will discuss three kinds of backgrounds: accidental, correlated and Cosmogenic Isotopes.
\subsubsection{Accidental Background}\label{sec:accidental}
The prompt component of the background signal is produced by natural radioactivity within the detector.  To reduce potential effects of this background, the modern experiments all utilize a buffer volume of non-scintillating oil to shield the active regions of the detector from the naturally-radioactive photomultipliers.  Also, emphasis has been placed on radio purity during the choice of detector materials and detector construction.  This background may typically be measured to high precision. This background will be considered in the rest of this review as fully correlated between detectors.
\subsubsection{Correlated Backgrounds}\label{sec:correlated}
A fast neutron is produced by nuclear spallation outside of the detector and interacts with a proton within the detector, causing a prompt signal as the proton recoils and scintillates.  The neutron then thermalizes within the detector, and captures to provide the delayed component of the signal. Most of the time the stopping-muon background category is included in the Correlated Background: a muon entering the detector interacting in the scintillator (prompt signal) and then stopping and releasing a Michel electron (delayed signal). For simplicity and because some details of the reactor experiment analysis are currently not well known, we will also consider this background as correlated between detectors but we would like to mention that this is not entirely correct: the different locations of the far and near detectors in terms of overburden determine different muon fluxes and a different muon energy spectrum. The muon flux determines the rate of production of long-lived isotopes like $^9$Li and $^8$He. To compensate for this effect, the far and near detector outer veto designs need to be modified as well as the signal selection cuts, which introduces additional systematic errors uncorrelated between detectors. 
\subsubsection{Cosmogenic Isotopes}\label{sec:cosmogenics}
A cosmic-ray muon interacts with a $^{12}$C nucleus and causes an inelastic scattering, creating long-lived neutron-heavy isotopes such as $^9$Li and $^8$He. These isotopes have mean lifetimes on the order of milliseconds, preventing their being vetoed.  They mimic an IBD signal when they do finally decay, with particles released in the decay providing the prompt component, and often releasing one or more neutrons to capture as a delayed component. The same considerations as for the previous background apply in this case.
%
%
\subsection{One detector analysis: Reactor spectrum and flux determination}\label{sec:reactor_spectrum}
The goal for a one detector analysis is the correct estimation of the \nuebar reactor flux. First ingredients is the measurement of the reactor thermal power for each of the reactor cores. Based on that measurements, and knowing the fuel composition of the reactors, the burn-up state is calculated as a function of time using one of the available reactor simulation codes (for example MURE or DRAGON\cite{DoubleChooz,takahama_paper}). A reactor simulation code produces an instantaneous fission rate for each of the reactor fissile isotopes as a function of time in the fuel burn up cycle. At this point the neutrino spectrum is calculated from the fission rate and as last step, the neutrino spectrum is converted into a positron spectrum. We will give a brief description here of these three steps. For a more detail discussion on how the reactor flux is obtained please refer to G. Gratta {\it et al.}\cite{Gratta_paper}.
A modern Pressurized Water Reactor (PWR) has a thermal power of 3~GW$_{th}$/core and the reactor fuel is enriched in $^{235}$U. $\sim$~6~$\bar{\nu_e}$ are produced by each reactor fission with a typical flux of  $\sim 6\times 10^{20}$~{$\bar{\nu_e}$}~core$^{-1}$~s$^{-1}$. The four main fissile isotopes are $^{235}$U, $^{238}$U, $^{239}$Pu, and $^{241}$Pu. Their fission rates are used as an input for the prediction of the $\bar{\nu}_e$ spectrum. In the past measurements of the $\beta$ spectrum for each of the fissile isotopes were done\cite{Schreckenbach,Hahn}. Two types of uncertainties are important at this stage: uncertainties in the initial fuel composition and measurement of the reactor parameters, generally performed by the power company that maintains and operates the reactors, that are given as input to the reactor core simulation code. Uncertainties due to simulation itself are also important, those simulations code are usually benchmarked using experimental reactor data a specific report could be found in C. Jones {\it et al.}\cite{takahama_paper}.
To determine the \nuebar spectra associated with the main three fissile isotope ($^238$U is at this stage excluded), the electron spectra need to be converted in positron spectrum.
$\frac{{\rm d}N}{{\rm d}E_e} \simeq \frac{{\rm d}N}{{\rm d}E_{\bar{\nu}_e}}$, where $E_e = \sqrt{p_e^2 + m_e^2}$ is the full electron energy. The electron spectrum equation is:
\begin{eqnarray}
Y(E_e)  =  \int_{E_e}^{\infty} {\rm d}E~n(E,Z)~k(E,Z)~p_e~E_e~(E - E_e)^2~F(E_e,Z)
\label{eq:spectrum}
\end{eqnarray}
where $n(E,Z)$ is a function describing the distribution of endpoints $E$ and nuclear charges $Z$, $k(E,Z)$ is the spectrum normalization constant, $p_e$ is the electron momentum, and $F(E_e,Z)$ is the Fermi function describing the Coulomb effect on the emitted electron. The endpoint function $n(E,Z)$ could be written as:
\begin{eqnarray}
n(E,Z) = - \frac{1}{2 k(E,Z)} \frac{{\rm d^3}}{{\rm d}E^3}
\left( \frac{Y(E)}{p E F(E,Z)} \right).
\label{eq:endpoint}
\end{eqnarray}
The expressions in Eq.~\ref{eq:spectrum} is too complicated to be used and so the integral in Eq.~\ref{eq:spectrum} is often replaced by a sum over the various beta decay branches. More details on how the $\bar{\nu_e}$ spectra is calculated can be found in Refs. (Schreckenbach {\it et al.}\cite{Schreckenbach}, Hahn {\it et al.}\cite{Hahn}, Davis  {\it et al.}\cite{Davis}, Vogel {\it et al.}\cite{Vogel:1981}, Klapdor and Metzinger\cite{Klapdor:1982a}, G.~Gratta {\it et al.}\cite{Gratta_paper} and P.~Huber {\it et al.}\cite{Huber:2003pm}.). A more analytical approximation is given in Vogel and Engel\cite{VogelEngel:1989}.
%
%
%
\paragraph{From \nuebar to positrons}\label{sec:nuebartopositron}
All reactor experiments (if not using a near detector) need to know the positron spectra once they determine the \nuebar spectra and this implies knowing the IBD cross section. The total cross section for IBD is defined as:\cite{Gratta_paper}.
\begin{eqnarray}
\sigma^{(0)}_{tot}  = 0.0952 \left(\frac{ E_e^{(0)} p_e^{(0)}}{1 {\rm\ MeV}^2}\right)
\times 10^{-42} {\rm\ cm}^2\,
\label{eq:sigtot0}
\end{eqnarray}
\noindent where $ E_e^{(0)} = E_{\nu} - (M_n - M_p) $ is the positron energy without considering the proton recoil correction, and $p_e^{(0)}$ is the momentum. 
\noindent The (small) energy-dependent outer radiative corrections to $\sigma_{tot}$ are given in Vogel\cite{Vogel:1984} and Fayans\cite{Fayans:1985}. Additional corrections to the cross section of order $E_{\nu}/M$ are described in Vogel {\it et al.}\cite{VogelBeacon:1999}. 
\par In order to check the goodness of the prediction results, in terms of $\bar{\nu}_e$ energy spectrum and total flux normalization, we need to use short baseline ($<$100~m) reactor oscillation experiment measurements.
Recent global fits\cite{mention_sens,mezzetto,Fogli} to all available short-baseline reactor antineutrino data\cite{bugey,bugey3,bugey4} have shown a consistent deficit of detected neutrinos with respect to flux predictions, interpreted by some to be indicative of mixing with a sterile neutrino\cite{anomaly} at $\Delta m_{\text{new}}^2 \sim 1.5$ eV values. This hypothesized mechanism of disappearance at very short baselines presents two challenges to the current generation of reactor experiments: providing any further constraints on the anomaly-driven oscillations; and conducting a study of $\tot$-driven oscillations in a manner which takes into account the anomaly-driven oscillations. Modern multiple-baseline experiments may address the first challenge with high-statistics data 
from their near detectors.  While the $\sim$100~m baselines of these detectors are too long to provide sensitivity to $\Delta m_{\text{new}}^2$, accurate measurements of the \nuebar flux before it experiences $\tot$-driven oscillations could provide additional information about the potential magnitude of the anomaly-driven mixing.  Data from Double Chooz's first results\cite{DoubleChooz} have already been found to lend additional support to the consistent deficit by C.~Giunti and M.~Lavender\cite{giunti_anomaly}.
Approaches to the second challenge vary by experimental configuration. In a single-detector experiment, \emph{a la} Double Chooz\cite{DoubleChooz}, a comparison of data from a baseline of $\sim$1~km to flux predictions is sensitive to the effects of anomaly-driven oscillations.  To mitigate these effects, a high-precision rate measurement (\emph{e.g.} at $\sim$40~m Y. Declais {\it et al.}\cite{bugey4}) may be used to ``anchor" the predicted flux to a normalization which already accounts for any anomaly-driven oscillations.  The experiment used to provide the rate measurement must have been at an appropriate baseline: far enough from the 
reactor to measuring a \nuebar flux which has already suffered anomaly-driven oscillations, but close enough to 
still be free of $\tot$-driven oscillations.  In a multiple-detector experiment, especially the current 
generation, the near detectors themselves will be situated at baselines where they can fill this role.

\subsection{Near Detector Data as a ``Predicted'' Flux}\label{sec:near_prediction}
The current generation of reactor antineutrino experiments will circumvent the uncertainties intrinsic to 
the calculation of the reactor antineutrino flux by the use of a near detector. In principle, a near detector measures 
the emitted spectrum at a baseline where $\theta_{13}$-driven oscillations are negligible. On the basis of statistics 
alone, the intrinsic uncertainty of the reactor flux prediction can be superseded with a measurement of as few as 3,000 events in a near detector. At their estimated near detector event rates, this can be achieved with only a few weeks of near detector data-taking by the three modern reactor experiments. However, since inter-detector systematics become the limiting factors in a multi-detector oscillation analysis, additional statistics are desirable. This is especially true in cases where the far detector, having a much lower signal event rate than the near, begins data taking at roughly the same time: the sensitivity to oscillations may gradually improve with time as the far detector accumulates data. 
As well-explained by P. Huber {\it et al.}\cite{Huber:2003pm}, when the far detector records a very large data set, then the normalization get set by the far detector data itself and becomes more accurate and reliable than any other inputs. This means that the sensitivity limit becomes nearly independent from the normalization, and we have a shape-only dependence (bin-to-bin correlation). At this stage, the inter-detectors energy calibration difference becomes very important.
%
%
\section{Physics Potential}\label{sec:potential}
\subsection{Calculation Parameters}\label{sec:parameters}
Attributes of each experiment's configuration, such as reactor and detector locations, detector target tonnage, 
reactor power, etc., have been taken from literature\cite{DayaBay,DoubleChooz,Reno} A summary of the principal systematic uncertainties and their presumed values can be found in Tab.~\ref{tab:common_unc}.  
Estimated detector efficiencies are taken from the experiments' proposals, with some modifications to reproduce expected signal event rates.  We presume that each detector has an 80\% duty cycle for physics data-taking.  Each detector's efficiency is presumed to be known to within 1.5\%, uncorrelated between detectors, with relative efficiencies between detectors known to the $\sim$0.5\% level. The energy scale of each detector is presumed to be linear and known to an uncertainty of $\sigma_{E-scale} = 7.5$\%, an optimistic estimate for first results, but achievable with multiple calibration campaigns. The \nuebar signal and backgrounds are conservatively treated as having independent, uncorrelated energy scales due to the different particle species producing each spectrum.
Three backgrounds are included in our study: accidental background events, with a falling exponential spectrum; fast neutron recoil events, implemented as a flat spectrum; and cosmogenic $^9$Li events, implemented using a nuclear decay Monte Carlo generator as the Double Chooz experiment\cite{DoubleChooz}. The rates of each background in each detector are taken from the experiments' proposals where available, from measurements in the case of Double Chooz\cite{DoubleChooz}, and from previous scalings accounting for overburden\cite{mention_sens}. A summary of the daily background rates used may be found in Tab.~\ref{tab:backgrounds}.  All background rates are presumed to be uncorrelated between detectors.  This presumption is valid for all cases except that of the cosmogenic backgrounds of Daya Bay, where multiple detectors occupying a single detector hall might be considered to share background rates.  In that case, our presumption is conservative.  The uncertainties on the background rates are rooted in what constraints are capable by performing an energy-dependent oscillation analysis where the backgrounds are allowed to vary as pull parameters. 
Each reactor is assumed to have a 91\% average uptime, consistent with being offline one month each year for fuel rod repositioning/refueling. The power of each reactor is considered to have an uncertainty of $\sigma_{pwr} = 1$\%, a value within range of those given in recent results from the Double Chooz experiment\cite{DoubleChooz} and those in the experiments' proposals. These power uncertainties are uncorrelated between reactors.  Also included are uncertainties on the fuel distributions of the reactors, implemented as four pull parameters per reactor with intra-reactor correlations, but no inter-reactor correlations. The cross-section per fission of the IBD interaction is presumed to have an uncertainty $\sigma_{xsec} = 2$\% that is correlated between reactors, including uncertainties from both the reference spectra and the normalization of the IBD cross-section. An increased $\sigma_{xsec}$ directly impacts the sensitivities of single-detector experiments, but would have no effect on multi-location experiments.  
\begin{table}[htbn!]
\tbl{List of used background rates, in events per day.}
{\begin{tabular}{|c|c|c|c|}
\toprule \hline
& Accidentals (d$^{-1}$) & Fast Neutrons (d$^{-1}$) & Cosmogenic $^9$Li (d$^{-1}$) \\ \hline
Double Chooz Near\protect\cite{mention_sens} & 13.6 $\pm$ 1.36 & 1.36 $\pm$ 0.68 & 9.52 $\pm$ 2.38 \\
Double Chooz Far\protect\cite{DoubleChooz} & 0.30 $\pm$ 0.03 & 0.82 $\pm$ 0.38 & 2.3 $\pm$ 1.2 \\ \hline
RENO Near\protect\cite{Reno,mention_sens} & 7.1 $\pm$ 0.71 & 3.0 $\pm$ 1.5 & 2.8 $\pm$ 0.7 \\
RENO Far\protect\cite{Reno,mention_sens} & 0.90 $\pm$ 0.09 & 1.0 $\pm$ 0.5 & 0.70 $\pm$ 0.18 \\ \hline
Daya Bay DYB\protect\cite{DayaBay} & 1.86 $\pm$ 0.19 & 0.5 $\pm$ 0.25 & 3.70 $\pm$ 0.93 \\ 
Daya Bay LA\protect\cite{DayaBay} & 1.52 $\pm$ 0.15 & 0.35 $\pm$ 0.18 & 2.50 $\pm$ 0.63 \\ %
Daya Bay Far\protect\cite{DayaBay} & 0.12 $\pm$ 0.01 & 0.03 $\pm$ 0.02 & 0.26 $\pm$ 0.07 \\ \hline
\bottomrule
\end{tabular}\label{tab:backgrounds} }
\end{table}
\begin{table}[ht!]
\tbl{List of used uncertainties.  All uncertainties have been implemented as pull parameters, 
with corresponding pull terms in the $\chi^2$ function. Treatment of the reactor uncertainty is done as in C.~Jones {\it al.}\protect\cite{takahama_paper}.}
{\begin{tabular}{|c|c|c|} \toprule \hline
 Uncertainty Type  &  Value  \\ \hline
Power (uncorr. b/w reactors) $\sigma_{pwr}$ & 1\% \\
Fuel Composition (uncorr. b/w reactors) $\sigma_{fuel}$ & $\sim$5\% \\ \hline
IBD Cross Section (corr. b/w det.) $\sigma_{xsec}$ & 2\% \\
Absolute Det. Eff. (uncorr. b/w det.) $\sigma_{det}$ & 1.5\% \\
Relative Det. Eff. (corr. b/w det.) $\sigma_{rel}$ & 0.38\%(DB) \\
& 0.5\% (DC) \\
& 0.6\% (RENO) \\
Signal Energy Scale (uncorr. b/w det.) $\sigma_{E-scale}$ & 7.5\%  \\
Background Energy Scale (uncorr. b/w det.) $\sigma_{B-scale}$ & 7.5\% \\
Background $i$ Rate (uncorr. b/w det.) $\sigma^i_{bkg}$ & (see Tab.~\ref{tab:backgrounds}) \\ \hline \bottomrule
\end{tabular}\label{tab:common_unc} }
\end{table}
%
%
\subsection{Presumed Experimental Schedules} \label{sec:timescales}
At the time of this writing, all three modern reactor experiments have begun taking data 
at different times and with different detector and site configurations. Double Chooz commenced taking data with its 
far detector in April 2011, and plans first-light for its near detector by early 2013.  
RENO began a calibration campaign in August 2011, with physics 
data-taking beginning with both near and far detectors in the following month.  Daya Bay began 
taking data with its DYB near detectors in August 2011.  Indications are that their far detectors 
will be operational by early spring 2012, with the Ling Ao detectors coming on-line in early summer 2012.  
We presume that each experiment will continue running for at least three years 
after its completion, in order to reach a sensitivity regime where statistical uncertainties 
are unimportant compared to systematics.
\begin{figure}[htbn!]
\begin{center}
\includegraphics[width=0.8\columnwidth]{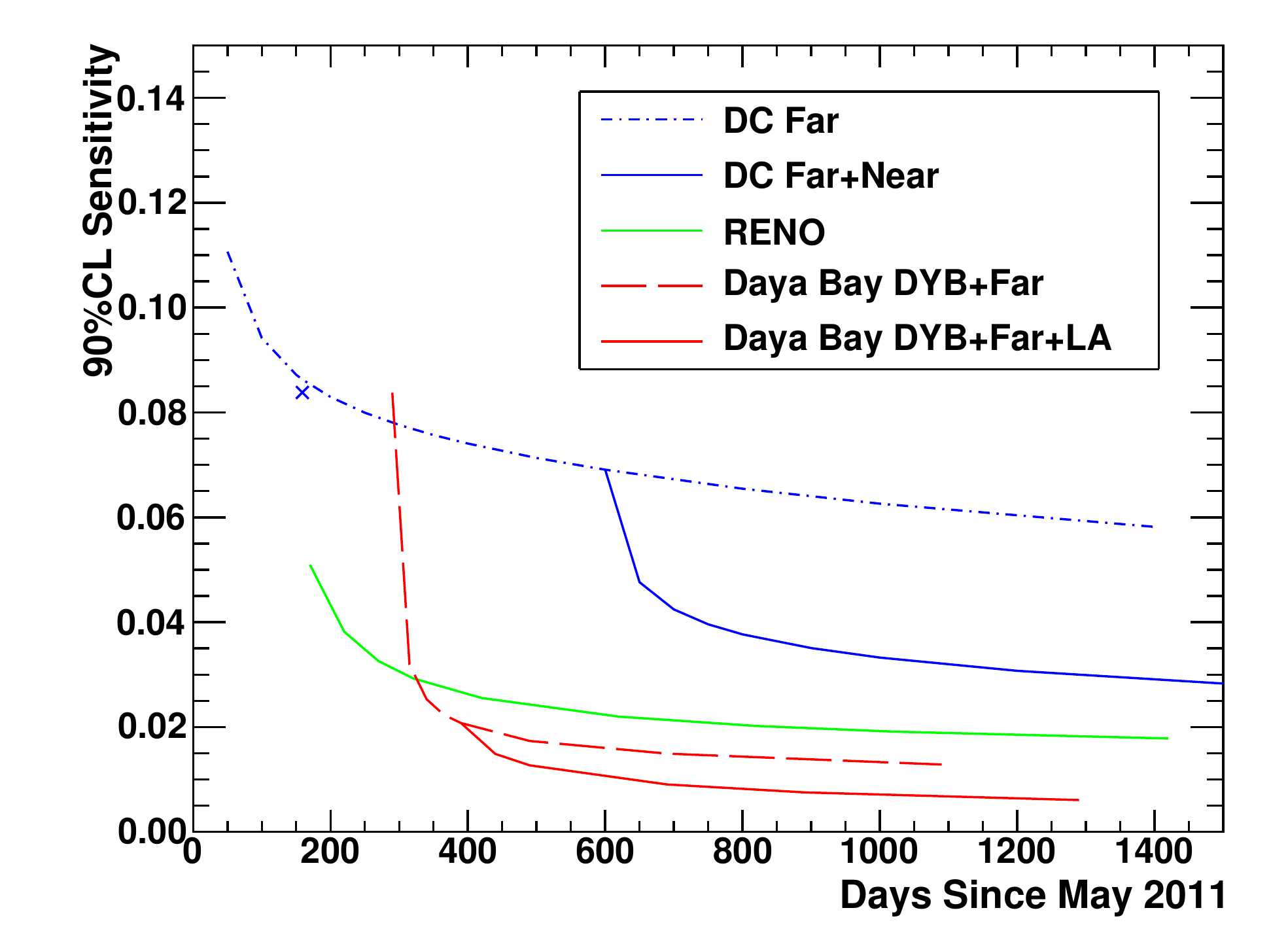}
\caption{Results of sensitivity calculations at 90\% CL based on projected experiment schedules described in Sec.~\ref{sec:timescales}. Presumed background rates are given in Tab.~\ref{tab:backgrounds}, and presumed common uncertainty parameters are shown in Tab.~\ref{tab:common_unc}.  The Double Chooz first result\protect\cite{DoubleChooz} is reported in the graph as well as sensitivity limit. The timescale for each experiments is computed considering a detector up-time of 80\% that includes a long calibration period with no data taking every 3 to 6 months.}
  \label{fig:time_sens}
\end{center} 
\end{figure}
%
%
\subsection{Single/Multi-Detectors Experiment} \label{sec:single_detector}
While all of the modern reactor \nuebar-disappearance experiments are designed around the concept 
of multiple detectors, some will take data while partially-constructed, potentially warranting a 
single-detector analysis. This has already occurred in the case of Double Chooz\cite{DoubleChooz}, 
and may occur as Daya Bay takes data with its ``DYB" near detectors. RENO has begun simultaneous operation 
of its two detectors, and will not undertake a single-detector phase.  
\par While running with a single detector, the dominant sources of uncertainty are the predicted \nuebar flux, the absolute detector efficiency, and background rate uncertainties. Statistical uncertainties may also be significant, due to the early-stage nature of the experiments when analyses of this type are performed. While the rates of the cosmogenic $^9$Li and fast neutron recoil backgrounds contribute significant uncertainty to a rate-only oscillation analysis, with sufficient statistics they may be constrained by data binned in energy. At energies around and above 8 MeV, the signal contribution is low enough such that the background contributions are dominant, allowing them to be constrained by the data themselves. Both Double Chooz and Daya Bay single-detector running may benefit from periods where one or more reactors are at reduced power or completely powered-down.  In the case of Double Chooz far-only running, this would permit \emph{in situ} measurements of background rates and spectra, similar to what the original CHOOZ was able to perform before the Chooz reactors became active\cite{Chooz}. The Double Chooz site, with its two reactors, is unique in providing this prospect: a higher number of nearby reactor cores decreases the probability that all might be shut down simultaneously. Daya Bay's single-detector efforts benefit from reactor-off time for different reasons: while the DYB detectors can be considered ``far" detectors to the Ling Ao and Ling Ao-II reactor cores, disappearance signals from those detectors' fluxes are overwhelmed by the flux from the nearly DYB reactors. A short (between 7 and 14 days) period of both of the DYB reactors in a low-power state could have the effect of improving their measurement sensitivity on $\sin^22\theta_{13}$ at 68\% CL after 200 days' data-taking from 0.051 to between 0.043 and 0.038, depending on the length of the ``off" period.
In the cases of RENO, Daya Bay and Double Chooz (near + far detector) uncertainties on the reactor flux are nearly nullified, and the dominant uncertainties are the relative detector efficiencies and the background rates. To reduce the relative detector efficiencies each of the reactor experiments uses different techniques and the background rates are constrained performing energy-dependent analysis. Let us consider the following hypothetical scenario as an example: both the far and near detectors of Double Chooz turn on simultaneously, and take data for 350 days. In this configuration, performing a rate-only analysis yields a sensitivity to $\sin^22\theta_{13}$ at 68\% CL of 0.070, while an analysis binned in energy, taking into account rate and spectral shape, yields a sensitivity of 0.036 at 68\% CL. In the latter case in particular, the cosmogenic background rates are constrained by data to better intervals than the \emph{a priori} uncertainty intervals. 
\par We have also studied the impact of the reactor flux uncertainty on the RENO experiment. The RENO near detector is at a distance ranging from 300 to 740~m from the reactor cores and most of the \nuebar flux comes from the 2 closest reactors at a distance of $\sim$350~m. We found a decrease on the sensitivity to $\sin^22\theta_{13}$ at 68\% CL of $\sim$ 4\% (from 0.027 to 0.028) while increasing the reactor flux uncertainties from 5\% to 15\%. We have also evaluated the sensitivity to $sin^22\theta_{13}$ for the Daya Bay experiment with only  ``DYB" site data considering 300 days of actual data taking. The ``DYB" site see most of the \nuebar flux coming from the DYB closest cores at 360 m while the rest of the flux is coming from 2 different baselines (1,1~km and 1,6~km respectively from Ling Ao-I and Ling Ao-II). We calculate a sensitivity for $\sin^22\theta_{13}$ at 68\% CL equal to $0.051$. The DYB sensitivity result shows the power of the rate+shape analysis and it is included just as an example.
\section{Conclusion}\label{sec:conclusion}
The use of nuclear reactors to study neutrino oscillations started with the CHOOZ\cite{CHOOZ} experiment and in recent years has reached a high level of sophistication. Actual reactor experiments use multiple detectors, multiple baselines and optimized designs to reject backgrounds. The use of multiple detectors make the uncertainties due to reactor flux calculations negligible but a lot of progress has been made also in the accuracy of the reactor neutrino flux and spectrum predictions. The latest result from Double Chooz of $sin^22\theta_{13} =$ 0.086 $\pm$ 0.041~({\rm stat}) $\pm$0.030~({\rm syst})\cite{DoubleChooz} has shown how a reactor experiment can achieve good sensitivity for $sin^22\theta_{13}$ without the use of a near detector. New results from Reno, Daya Bay and Double Chooz are expected in 2012. 
\end{document}